\documentclass[apjl]{emulateapj} 

\usepackage{epsfig}
\usepackage{amsmath}
\usepackage{amssymb}
\usepackage{natbib}
\usepackage{threeparttable} 
\usepackage{booktabs}

\usepackage{epstopdf}

\newcommand{\chan}{\textit{Chandra}}
\newcommand{\swift}{\textit{Swift}}

\newcommand{\xmm}{\textit{XMM-Newton}}

\newcommand{\maxi}{\textit{MAXI}}

\newcommand{\Msun}{\mathrm{M}_{\odot}}
\newcommand{\lum}{\mathrm{erg~s}^{-1}}
\newcommand{\flux}{\mathrm{erg~cm}^{-2}~\mathrm{s}^{-1}}

\newcommand{\cnts}{\mathrm{counts~s}^{-1}}

\newcommand{\nh}{\mathrm{cm}^{-2}}
\newcommand{\y}{\mathrm{g~cm}^{-2}}
\newcommand{\dist}{(D/8.5~\mathrm{kpc})^2}

\newcommand{\source}{XTE~J1709--267}

\newcommand{\rosatname}{1RXS J170930.2--263927}

\slugcomment{To appear in the Astrophysical Journal Letters}

\shorttitle{\source\ in quiescence}
\shortauthors{Degenaar, Wijnands, \& Miller}

\begin{document}


\title{A direct measurement of the heat release in the outer crust of the transiently accreting neutron star \source}

\author{N. Degenaar$^{1,}$\altaffilmark{3}, R. Wijnands$^{2}$, and J. M. Miller$^{1}$}
\affil{$^1$Department of Astronomy, University of Michigan, 500 Church Street, Ann Arbor, MI 48109, USA; degenaar@umich.edu\\
$^2$Astronomical Institute Anton Pannekoek, University of Amsterdam, Postbus 94249, 1090 GE Amsterdam, The Netherlands\\}

\altaffiltext{3}{Hubble fellow}


\begin{abstract}
The heating and cooling of transiently accreting neutron stars provides a powerful probe of the structure and composition of their crust. Observations of superbursts and cooling of accretion-heated neutron stars require more heat release than is accounted for in current models. Obtaining firm constraints on the depth and magnitude of this extra heat is challenging and therefore its origin remains uncertain. We report on \swift\ and \xmm\ observations of the transient neutron star low-mass X-ray binary \source, which were made in 2012 September--October when it transitioned to quiescence after a $\simeq$10 week long accretion outburst. The source is detected with \xmm\ at a 0.5--10 keV luminosity of $L_{\mathrm{X}} \simeq 2 \times 10^{34}~\dist~\lum$. The X-ray spectrum consists of a thermal component that fits to a neutron star atmosphere model and a non-thermal emission tail, which each contribute $\simeq$50\% to the total flux. The neutron star temperature decreases from $\simeq$158 to $\simeq$152~eV during the $\simeq$8 hr long observation. This can be interpreted as cooling of a crustal layer located at a column density of $y\simeq5\times10^{12}~\y$ ($\simeq$50~m inside the neutron star), which is just below the ignition depth of superbursts. The required heat generation in the layers on top would be $\simeq$0.06--0.13~MeV per accreted nucleon. The magnitude and depth rule out electron captures and nuclear fusion reactions as the heat source, but it may be accounted for by chemical separation of light and heavy nuclei. Low-level accretion  offers an alternative explanation for the observed variability.
\end{abstract}



\section{Introduction}\label{sec:intro}
Many observable properties of neutron stars are set by the structure and composition of their crust. Low-mass X-ray binaries (LMXBs) offer powerful means to investigate the properties of these layers. In LMXBs, a neutron star accretes matter from a late-type companion star that overflows its Roche lobe. This typically generates a 2--10 keV luminosity of $L_{\mathrm{X}} \simeq 10^{35-38}~\lum$. Transient LMXBs only occasionally experience outbursts of accretion, while they spend the majority of their time in a quiescent phase with little or no matter being accreted onto the neutron star. The quiescent X-ray emission is therefore much dimmer; $L_{\mathrm{X}} \simeq 10^{31-34}~\lum$. 

Matter that falls onto the surface of a neutron star is compressed to higher densities as it becomes buried by freshly accreted material. As a result, the matter undergoes compositional changes that are driven by nuclear reactions. These involve thermonuclear burning near the stellar surface, electron captures and neutron emissions that occur a few tens of meters inside the neutron star, and nuclear fusion reactions that take place at several hundred meters depth \citep[][]{sato1979,wallace1981,haensel1990a,schatz2001}.

Initial calculations suggested that this chain of nuclear processes deposits an energy of $\simeq$1--1.5~MeV per accreted nucleon in the neutron star crust \citep[][]{haensel1990a,haensel2003}. However, observations of superbursts and the thermal evolution of neutron stars after an accretion phase both require a hotter crust than these heating rates account for \citep[][]{cumming06,brown08,degenaar2011_terzan5_3}. The discrepancy may be resolved if there is an additional source of heat from nuclear reactions or other processes in the outer crustal layers \citep[][]{gupta07,horowitz2008,medin2011}.

During outburst, the energy that is generated in the crust creates a temperature profile that depends on the depth and magnitude of the heat sources. During quiescence, the gained heat is thermally conducted toward the surface and the core, until the crust has cooled and thermal equilibrium is restored. The cooling of neutron star crusts can be observed as a decrease in quiescent thermal X-ray emission after the end of an accretion outburst \citep[][]{wijnands2002,wijnands2004,cackett2008,cackett2010,degenaar09_exo1,degenaar2011_terzan5_3,degenaar2010_exo2,diaztrigo2011,fridriksson2010,fridriksson2011}. 

Comparing crust cooling observations with neutron star thermal evolution models yields important insight into the properties of the crust and core \citep[][]{rutledge2002,shternin07,brown08, degenaar2011_terzan5_3,page2012}. However, the current data provide only weak constraints on any additional heat sources. Progress can be made by observing the temperature evolution shortly (within $\simeq$2~weeks) after the end of an outburst, since that can provide a direct measure of the depth and magnitude of the heat release in the outer crust \citep[][]{brown08,degenaar2011_terzan5_3}.

 \begin{figure*}
 \begin{center}
	\includegraphics[width=12.0cm]{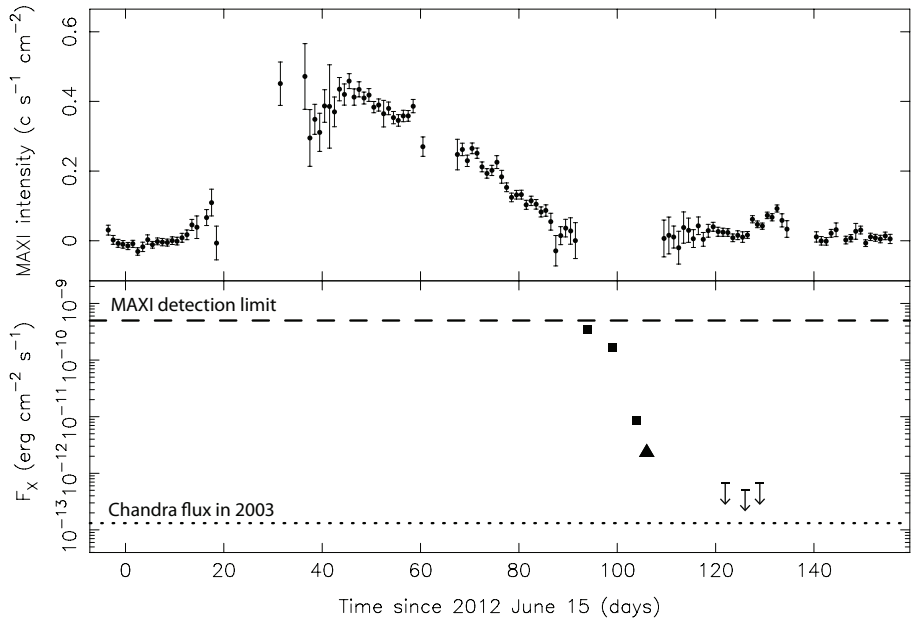}
    \end{center}
    \caption[]{Top: one day averaged \maxi\ light curve of the 2012 outburst (2--20 keV). Bottom: evolution of the unabsorbed 0.5--10 keV flux after the source intensity dropped below the detection limit of \maxi\ (dashed line). Squares and upper limits represent \swift\ data, and the triangle indicates the \xmm\ observation. The dotted line corresponds to the flux measured with \chan\ in 2003.}
 \label{fig:maxi}
\end{figure*} 


\source\ (\rosatname) is a neutron star LMXB that was first reported in outburst in 1997 \citep[][]{marshall1997}. The source is thought to be associated with the globular cluster NGC 6293 \citep[][]{jonker2004}, which would place it at a distance of $\simeq$8.5~kpc \citep[][]{harris1996,lee2006}. In the past 15 years, it has exhibited several accretion outbursts with a typical 2--10 keV intensity of $L_{\mathrm{X}}\simeq2\times10^{37}~\dist~\lum$ \citep[][]{marshall1997,cocchi1998,jonker2003,jonker2004_atel,jonker2004,markwardt2004,remillard2007, negoro2010,yamauchi2010}. 

The most recent outburst occurred in 2012 June--September \citep[][]{sanchez2012}. Figure~\ref{fig:maxi} displays the light curve as observed with \maxi. The source was active for $\simeq$10~weeks with an average 2--10 keV flux of $F_{\mathrm{X}} \simeq 2 \times 10^{-9}~\flux$. Its intensity remained below the detection limit of \maxi\ from September 11 onward, marking its return to quiescence (Figure~\ref{fig:maxi}).

In this work we report on a series of \swift\ pointings and an \xmm\ observation of \source. The data were obtained in 2012 September--October, and aimed to search for signs of an accretion-heated crust.

\begin{table}
\begin{center}
\caption{X-Ray Observation Log\label{tab:obs}}
\begin{tabular*}{0.48\textwidth}{@{\extracolsep{\fill}}cccc}
\hline
\hline
Observatory &  Date & Obs ID & Exposure Time  \\
 &   &  & (ks)  \\
\hline
\swift & 2012 Sep 18 & 35717011 & 0.9  \\	
\swift & 2012 Sep 23 & 35717012 & 0.8   \\ 
\swift & 2012 Sep 28 & 35717013 & 0.9   \\  
\xmm & 2012 Sep 30 & 0700381401 & 31   \\
\swift & 2012 Oct 16 & 35717014 & 0.9   \\	
\swift & 2012 Oct 20 & 35717015 & 1.1   \\	
\swift & 2012 Oct 23 & 35717016 & 0.9   \\	
\hline
\end{tabular*}
\tablenotes{
{\bf Notes.} After excluding a short episode of background flaring, the net exposure time of the \xmm\ observation was $\simeq$30 ks for the PN and $\simeq$28 ks for the MOS1.
}
\end{center}
~\\
\end{table}


\section{Observations, Data Analysis and Results}
The observations discussed in this work are listed in Table~\ref{tab:obs}. All spectral data were fitted between 0.5 and 10 keV using \textsc{XSpec} \citep[version 12.7;][]{xspec}. Interstellar absorption was taken into account by using the \textsc{tbabs} model \citep[][]{wilms2000}. Throughout this work we assumed a source distance of $D=8.5$~kpc. All quoted errors refer to 1$\sigma$ confidence intervals.


\subsection{\swift}\label{subsec:swift}
\source\ was observed with \swift\ six times (Table~\ref{tab:obs}). Data reduction and analysis was carried out within \textsc{heasoft} (version 6.11). We re-processed the raw data obtained with the X-ray Telescope (XRT) using the \textsc{xrtpipeline}. Spectra and light curves were extracted using \textsc{XSelect}. 

The first two XRT observations were affected by pile-up. We therefore extracted source events using an annular region with inner and outer radii of $12''$ and $60''$ for observation 35717011, and $10''$ and $60''$ for 35717012. For the other four observations we used a circular region with a radius of $15''$. Background events were collected from a source-free circular region with a radius of $45''$.

\subsubsection{X-ray Spectra and Light Curve}\label{subsubsec:xrt}
The first series of \swift\ observations were obtained on September 18, 23 and 28. We fitted the XRT spectra of these three data sets simultaneously to a simple absorbed power-law model. This resulted in a joint value of $N_{\mathrm{H}} = (3.4\pm0.3) \times 10^{21}~\nh$ and power-law indices of $\Gamma=2.0\pm0.1$, $2.4\pm0.1$, and $2.6\pm0.3$ for the first, second and third observation, respectively (yielding $\chi^2_{\nu}=0.93$ for 118 dof). The corresponding unabsorbed 0.5--10 keV fluxes are $F_{\mathrm{X}} = (3.4\pm0.2) \times10^{-10}$, $(1.7\pm0.2) \times10^{-10}$ and $(8.4\pm1.3) \times10^{-12}~\flux$ (Figure~\ref{fig:maxi}). 

The intensity and spectral shape suggest that the source was still in outburst during the first two \swift\ observations, but that accretion may had ceased by the time of the third. We therefore attempted to fit this spectrum with the model inferred from the \xmm\ data (Section~\ref{subsubsec:xmmspec}). Leaving only the neutron star temperature and power-law normalization as free fit parameters results in $kT^{\infty} \gtrsim 191$~eV. This is not unphysical and may indicate that the neutron star surface was visible.

The second series of \swift\ pointings were performed on October 16, 20 and 23. \source\ is not detected in these observations. We determined 95\% upper limits on the count rate by applying the prescription for small numbers of counts given by \citet{gehrels1986}. Corresponding upper limits on the 0.5--10 keV unabsorbed flux were estimated using \textsc{pimms} (version 4.6) by assuming a power-law spectrum with $N_{\mathrm{H}} = 4.9 \times 10^{21}~\nh$ and $\Gamma = 2.9$ (see Section~\ref{subsubsec:xmmspec}). This yielded $F_{\mathrm{X}} \lesssim (3-7) \times10^{-13}~\flux$. Summing the three observations reveals a small excess of photons at the source position for which we estimate $F_{\mathrm{X}} \simeq 2 \times10^{-13}~\flux$. This is a factor of $\simeq$2 higher than the quiescent level observed with \chan\ in 2002 (see Section~\ref{subsubsec:xmmspec}). Figure~\ref{fig:maxi} displays the flux light curve of the \swift\ observations.


\subsection{\xmm}
\xmm\ observed \source\ for 31 ks starting on UT 2012 September 30 at 09:47. We analyzed the data obtained with the European Photon Imaging Camera (EPIC). 
The PN and the MOS1 were both set in small window imaging mode. The source was not detected with the MOS2, which was operated in timing mode. Data reduction and analysis was carried out using the Science Analysis Software (\textsc{SAS}; version 11.0.0). 

A short episode of background flaring was excluded by selecting data with high-energy count rates of $<$0.15$~\cnts$ for the PN (10--12 keV) and $<$0.30$~\cnts$ for the MOS1 ($>$10~keV). This resulted in a total net exposure time of $\simeq$30 and $\simeq$28~ks for the PN and the MOS1, respectively. We extracted source events from a circular region with a radius of $30''$, and used a source-free circular region with a radius of $60''$ for the background. 

\source\ is detected at an average net count rate of $0.568 \pm 0.006~\cnts$ with the PN and at $0.195 \pm 0.003~\cnts$ with the MOS1. Background-corrected light curves were created using the tasks {\sc evselect} and {\sc lccorr}. The X-ray spectra and response files were extracted using the task {\sc especget}. The spectral data were grouped to contain $>20$ photons per bin.

 \begin{figure}
 \begin{center}
	\includegraphics[width=8.7cm]{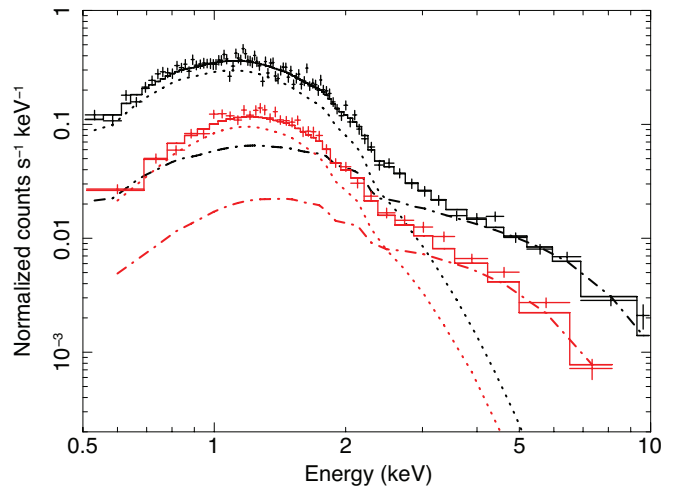}
    \end{center}
    \caption[]{PN (top, black) and MOS1 (bottom, red) spectra. The solid line represents the best fit to a combined neutron star atmosphere (dotted lines) and power-law (dashed-dotted curves) model. The spectra were rebinned for representation purposes.}
 \label{fig:spec}
\end{figure}

\subsubsection{Average X-ray Spectrum}\label{subsubsec:xmmspec}
We fitted the PN and MOS1 spectral data simultaneously with all parameters tied between the two instruments. A simple absorbed power-law model (\textsc{pegpwrlw}) yields $N_{\mathrm{H}}=(4.9\pm0.1)\times10^{21}~\nh$ and $\Gamma=2.90\pm0.04$ ($\chi^2_{\nu}=1.32$ for 506 dof). The fit can be significantly improved by adding a thermal component. 
We used the neutron star atmosphere model \textsc{nsatmos} \citep[][]{heinke2007}. In all fits we fixed the mass and radius to canonical values of $M=1.4~\Msun$ and $R=10$~km, adopted a distance of $D=8.5$~kpc, and set the normalization constant at 1 (i.e., the entire surface is radiating). This left the neutron star temperature as the only free fit parameter. 

The composite model provides a good fit, yielding $N_{\mathrm{H}}=(2.9\pm0.1)\times10^{21}~\nh$, $\Gamma=1.34\pm0.11$, and $kT^{\infty}=155.7\pm1.4$~eV  ($\chi^2_{\nu}=1.04$ for 505 dof). The spectral data and model fit are shown in Figure~\ref{fig:spec}. The inferred 0.5--10 keV unabsorbed flux is $F_{\mathrm{X}} = (2.33\pm0.01)\times10^{-12}~\flux$, which translates into a luminosity of $L_{\mathrm{X}} = (2.01\pm0.01)\times10^{34}~\dist~\lum$. The power-law component contributes $46\% \pm2$\% to the total unabsorbed flux. By setting the power-law normalization to zero and extrapolating the \textsc{nsatmos} model fit to the 0.01--100 keV range, we estimate a thermal bolometric flux of $F_{\mathrm{bol}} = (1.5\pm0.1)\times10^{-12}~\flux$. 

We compare these results with the quiescent properties measured with \chan\ on 2003 May 12 \citep[Obs ID 3507; for details, see][]{jonker2004}. We reduced the data using the \textsc{ciao} tools (version 4.4) and extracted a background-corrected spectrum. We fitted this with a combined power law and neutron star atmosphere model, with $N_{\mathrm{H}}=2.9\times10^{21}~\nh$ and $\Gamma=1.3$ fixed to the values inferred from the \xmm\ data. This yielded $kT^{\infty} = 88.6 \pm 8.7$~eV and a 0.5--10 keV unabsorbed flux of $F_{\mathrm{X}} = (1.32\pm0.09)\times10^{-13}~\flux$, of which $25\% \pm10$\% can be attributed to the power-law emission ($\chi^2_{\nu}=0.84$ for 14 dof).

 \begin{figure*}
 \begin{center}
	\includegraphics[width=12.0cm]{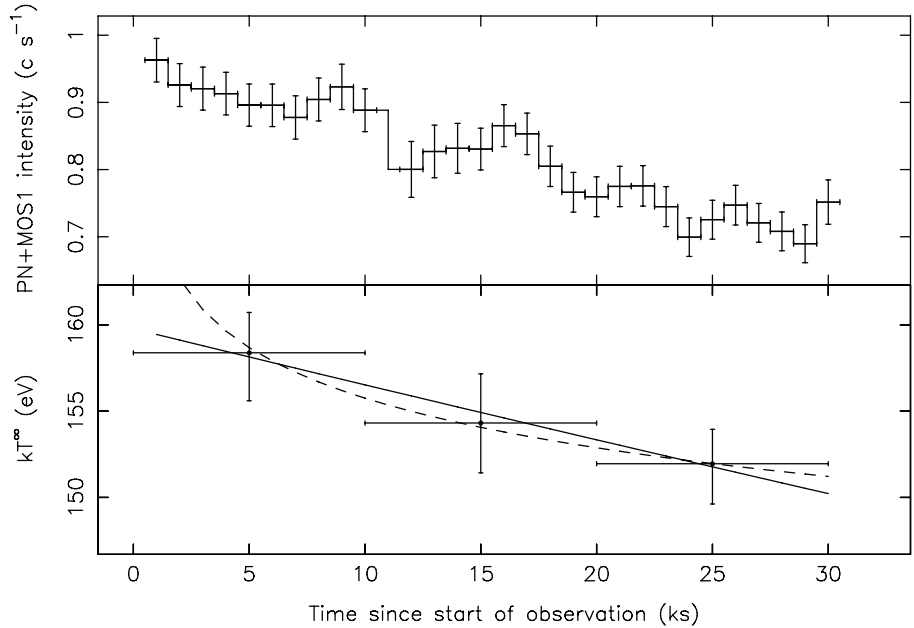}
    \caption[]{Top: combined PN/MOS1 light curve using a bin time of 1000 s. Bottom: evolution in neutron star temperature along with a fit to a power-law decay with an index of --0.03 (dashed curve) and an exponential decay of 5.6~days (solid line). 
     \label{fig:xmmlc}
    }
        \end{center}
\end{figure*} 

 \begin{figure}
 \begin{center}
	\includegraphics[width=8.7cm]{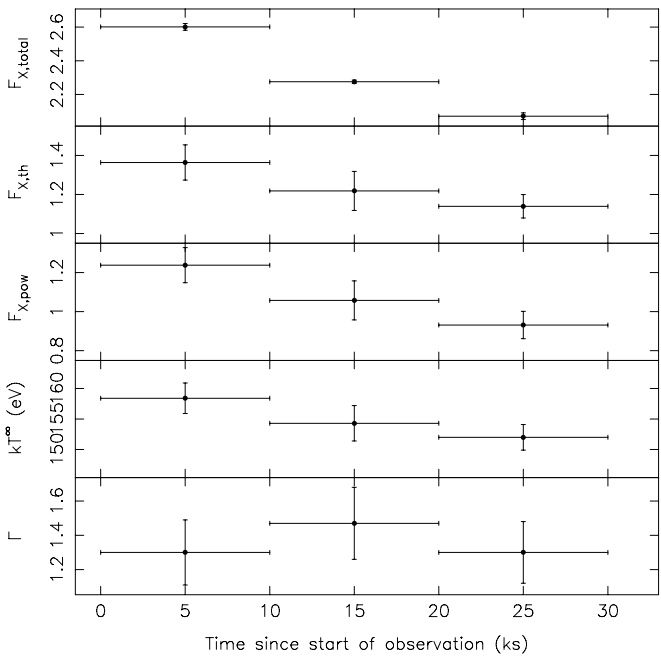}
    \end{center}
    \caption[]{Results of time-resolved spectroscopy of the PN and MOS1 data. From top to bottom: the total unabsorbed model flux, thermal flux, and power-law flux (0.5--10 keV and units of $10^{-12}~\flux$), the neutron star temperature, and the power-law index.}
 \label{fig:timeresolved}
\end{figure}

\subsubsection{A Decay in X-Ray Flux}\label{subsubsec:timespec}
The PN and MOS1 light curves reveal that the intensity of \source\ decreased during the observation. The co-added light curve is shown in Figure~\ref{fig:xmmlc}. To investigate the cause of this decay, we divided the data into three equal parts of $\simeq10$~ks and analyzed the individual spectra. We used the combined power law and neutron star atmosphere model described in Section~\ref{subsubsec:xmmspec} and required the hydrogen column density to be the same between the three data segments \citep[for a justification, see][]{miller2009}. 

The results of the time-resolved spectroscopy are shown in Figure~\ref{fig:timeresolved}. This suggests that the flux of both the thermal and the power-law component decreased during the observation. The fractional contribution of the power law to the total unabsorbed 0.5--10 keV flux was consistent with being constant: $48\% \pm3\%$, $46\% \pm3\%$ and $45\% \pm3\%$ for the first, second and third intervals, respectively. The neutron star temperature decreased $\simeq$2.5$\sigma$ during the observation, from $kT^{\infty} =158.4\pm2.5$ to $152.0\pm2.1$~eV. The power-law index remained constant within the errors (Figure~\ref{fig:timeresolved}).

~\\
\section{Discussion}\label{sec:discussion}
We report on \swift\ and \xmm\ observations obtained during the decay of the 2012 outburst of \source. The data cover a time span of 35 days, during which the source intensity decreases from $L_{\mathrm{X}} \simeq 3 \times 10^{36}$ to $\simeq 2 \times 10^{33}~\dist~\lum$ (0.5--10 keV). 

The spectral data can be described by an absorbed power law that softens from $\Gamma = 2.0$ to 2.6 in the first set of \swift\ observations, and further to $\Gamma=2.9$ when fitting the \xmm\ data with the same model. Similar softening has been observed in other transient LMXBs for which high-quality spectral data were obtained during the transition from outburst to quiescence \citep[for an in depth discussion, see][]{armas2011,armas2012}. The \xmm\ data of \source\ are much better described by adding a thermal component, which suggests that at least part of the observed softening is not caused by the power-law emission itself, but by a soft component becoming apparent in the spectrum.

\source\ is detected with \xmm\ at a 0.5--10 keV luminosity of  $L_{\mathrm{X}} \simeq 2 \times 10^{34}~\dist~\lum$. The spectral data are best fit with a composite model consisting of a thermal emission component that can be described by a neutron star atmosphere with a temperature of $kT^{\infty} \simeq 156$~eV, and a non-thermal emission tail that fits to a power law with an index of $\Gamma \simeq 1.3$ and contributes $\simeq$50\% to the total unabsorbed 0.5--10 keV flux. Such a spectral shape is common for neutron star LMXBs in quiescence. 

The 0.5--10 keV flux measured during the \xmm\ observation is a factor $\simeq20$ higher than seen with \chan\ in 2003 ($\simeq1$ year after the end of its 2002 outburst), and the inferred neutron star temperature is higher by a factor of $\simeq1.7$. The subsequent \swift\ observations show that the thermal emission had decreased, but was still a factor of $\simeq$2 above the level detected with \chan. This may indicate that (part of) the crust became substantially heated during the recent $\simeq10$-week outburst, and was cooling in quiescence. 

Most remarkably, we found that the intensity of \source\ decreased during the $\simeq$8-hr long \xmm\ observation. Time-resolved spectroscopy suggests that the neutron star temperature decreased from $\simeq$158 to $\simeq$152~eV. An attractive explanation is that the outburst indeed heated the crust and that it was cooling during our observation (see Section~\ref{subsec:heatflux}). However, the presence of a strong and variable non-thermal emission component suggests that residual accretion was possibly occurring. 

Multi-epoch studies of a number of neutron stars LMXBs have shown (non-monotonic) variations in the quiescent thermal emission that cannot be attributed to cooling of a heated crust \citep[][]{rutledge2002_aqlX1,campana2004,cackett2010_cenx4,fridriksson2010}. Correlated variability with the power-law component suggests that a residual accretion flow may reach the neutron star in some sources \citep[][]{cackett2010_cenx4,fridriksson2010}. This would generate a thermal spectrum that is difficult to distinguish from that of a cooling neutron star \citep[][]{zampieri1995,soria2011}. In this case the inferred temperature does not reflect that of the neutron star crust, but rather that of the surface that continues to be heated by the ongoing accretion.

Time-resolved spectroscopy indicates that the thermal and the power-law component may both have varied during the \xmm\ observation of \source. This suggests that matter was possibly still being accreted onto the neutron star. An alternative explanation is therefore that the observed flux decrease was caused by variations in the accretion rate.


\subsection{Constraining the Heat Release in the Outer Crust}\label{subsec:heatflux}
If the decrease in temperature observed during our \xmm\ observation was due to cooling of the neutron star crust, then we can use this result to constrain the heat release in the crust. At any given depth inside the neutron star it takes a certain characteristic time for heat to diffuse to the surface, i.e., for the layer to cool. The observed decrease in neutron star temperature from $kT^{\infty} \simeq 158$ to $\simeq152$~eV can be fitted with an exponential decay time of $\simeq5.6\pm1.9$~days (Figure~\ref{fig:xmmlc}). Using Equation (9) of \citet{brown08}, we find that this is the characteristic cooling time  for a layer located at a column depth of $y \simeq 5 \times 10^{12}~\mathrm{g~cm}^{-2}$ (corresponding to a matter density of $\rho \simeq 2\times 10^{9}~\mathrm{g~cm}^{-3}$, or a depth of $\simeq$50~m inside the neutron star). This is just below the depth where superbursts are thought to ignite \citep[][]{cumming06}. 

The decrease in temperature can also be described by a power-law decay with a slope of $-0.03 \pm 0.01$ (Figure~\ref{fig:xmmlc}). This can be used to estimate the heat flux in the crust; using Equations (8)--(11) of \citet{brown08}, we find $F_{\mathrm{crust}} = (1.0-2.0) \times 10^{21}~\flux$. Combining this with the mass-accretion rate provides an estimate of the energy that was released in the crust during outburst.

The average intensity observed with \maxi\ during the 2012 outburst translates into a 2--10 keV luminosity of $L_{\mathrm{X}} \simeq 2 \times 10^{37}~\dist~\lum$. The bolometric luminosity of neutron star LMXBs is typically a factor of $\simeq2.5$ higher than measured in the 2--10 keV band \citep[][]{zand07}, which would imply an average accretion luminosity of $L_{\mathrm{acc}} \simeq 4 \times 10^{37}~\dist~\lum$. This gives an estimated mass-accretion rate of $\dot{M}_{\mathrm{ob}}=RL_{\mathrm{acc}}/GM \simeq 2.1\times10^{17}~\mathrm{g~s}^{-1}$ (for $R=10$~km and $M=1.4~\Msun$). Assuming that the emission is isotropic, the mass-accretion rate per surface area is $\dot{m}_{\mathrm{ob}} \simeq 1.7 \times 10^{4}~\mathrm{g~cm}^{-2}~\mathrm{s}^{-1}$. We then find a heat release of $F_{\mathrm{crust}}/\dot{m}_{\mathrm{ob}}\simeq (6-12)\times10^{16}~\mathrm{erg~g}^{-1} \simeq 0.06-0.13$~MeV per accreted nucleon. This heat must have been produced in the layers above the cooling layer that is probed by our observation, i.e., at a depth of  $y \lesssim 5 \times 10^{12}~\mathrm{g~cm}^{-2}$.

We can compare these constraints with the different mechanisms that have been put forward as possible sources of extra heat release in the neutron star crust. The shallow depth ($y \lesssim 5 \times 10^{12}~\mathrm{g~cm}^{-2}$) rules out nuclear fusion reactions, since these occur much deeper in the crust \citep[$y \gtrsim 10^{14}~\mathrm{g~cm}^{-2}$;][]{horowitz2008}. Electron captures do occur at the inferred depth, but these are expected to generate a heat of $\simeq0.01$~MeV~nucleon$^{-1}$ \citep[][]{gupta07,estrade2011}, which is a factor of $\simeq$10 lower than we obtain. 

Chemical separation in the crust provides an alternative mechanism for the heat release. Just below the surface, the crust is composed of a variety of ions that effectively behave as a liquid. This is commonly referred to as the neutron star ``ocean'' and is thought to be the ignition site of superbursts. The matter freezes out at the bottom of the ocean, at $y\simeq10^{12}~\y$, becoming part of the solid crust. Molecular dynamics simulations suggest that the freezing of a mixture of elements causes chemical separation, with light nuclei mainly being left in the liquid phase and heavier nuclei preferentially becoming incorporated in the solid crust \citep[][]{horowitz2007}. The retention of light elements in the liquid ocean drives a buoyancy that can cause a convective heat flux of $\simeq0.01-0.2$~MeV per accreted nucleon, depending on the  composition \citep[][]{medin2011}. 

We conclude that within our current knowledge, electron captures and nuclear fusion reactions cannot be the source of shallow heat that may be present in \source. Instead, chemical separation of light and heavy nuclei can account for both the depth and the magnitude of the heat source that we infer from our \xmm\ observation. If cooling of a crustal layer is causing the observed flux decrease, this suggests that chemical separation may play an important role in heating the crusts of neutron stars. This process can then potentially supply the extra heat that is required to resolve the discrepancy between current heating models, observations of superbursts and crust cooling of accretion-heated neutron stars.


\acknowledgments
N.D. is supported by NASA through Hubble Postdoctoral Fellowship grant number HST-HF-51287.01-A from the Space Telescope Science Institute, which is operated by the Association of Universities for Research in Astronomy, Incorporated, under NASA contract NAS5-26555. 
R.W. is supported by a European Research Council starting grant. We are grateful to Norbert Schartel, Neil Gehrels, and the \xmm\ and \swift\ duty scientists for making these Target of Opportunity observations possible. This research made use of the data provided by RIKEN, JAXA and the MAXI team. 

{\it Facilities:} \facility{XMM (EPIC)}, \facility{Swift (XRT), \facility{MAXI}}

\end{document}